\documentclass[twocolumn,showpacs,amsmath,amssymb]{revtex4}

\usepackage{graphicx}
\usepackage{dcolumn}
\usepackage{bm}

\begin{document}


\title{Suppression of Landau damping  via electron band gap}

\author{S. Son}
\affiliation{18 Caleb Lane, Princeton, NJ 08540, USA}
\author{S. Ku}
\affiliation{Courant Institute of Mathematical Sciences, New York University, New York, NY 10012, USA}
\email{sku@cims.nyu.edu}

\date{\today}

\begin{abstract}

The pondermotive potential in the X-ray Raman compression can generate
an electron band gap which suppresses the Landau damping.
The regime is identified  where a Langmuir wave can be driven without damping
in the stimulated Raman compression. 
It is shown  that the partial wave breaking and the frequency detuning due to
the trapped particles would be greatly reduced.
\end{abstract}

\maketitle

Coherent intense X-rays might be feasible in near future thanks to
the advances in the fields of free electron laser~\cite{Free,Free2} and 
the inertial confinement fusion~\cite{lindl,tabak}.
A short coherent X-ray laser of durations of femto-seconds, would find
many applications~\cite{Protein,Protein2,fast2,fast4,fast5}.
Even shorter and more intense X-rays would enable probing and manipulating
small scale physical processes of ultra-fast time scales.
One promising approach for such an ultra short light pulse (even to a few atto-seconds~\cite{Fisch,Fisch2})
is the backward Raman scattering (BRS), where a pulse gets compressed via laser-plasma interactions. 
The BRS has already been used to compress the visible light~\cite{Ping}.
Recent theoretical analysis has attempted to examine the plausibility of this method
in the X-ray regime~\cite{Fisch,Fisch2}.

We note that some key physical processes of the X-ray BRS compression in
dense plasmas are  considerably different from those of the visible light.
In particular, due to the short wave length of  the  pondermotive potential,
the quantum diffraction and the degeneracy of the electron states become
relevant to the Landau damping and the frequency detuning~\cite{Oneil,Oneil2}.
In this letter, we show  that a sufficiently intense wave modifies
the electron's momentum energy dispersion relation and forms an electron band
structure, by which the Landau damping gets suppressed~\cite{Landau}. 
We also show that the wave breaking via the trapped electrons gets considerably reduced.


Consider a wave in the form of $\phi(x,t) = e\phi_0 \cos(kx -\omega t)$.
The classical analysis of the Landau damping rate, $\gamma_{\mathrm{cl}}$, for small $e\phi_0$ shows that~\cite{Landau} 

\begin{equation}
 \frac{\gamma_{\mathrm{cl}}}{\omega} = \frac{\pi}{2} \frac{\omega_{\mathrm{pe}}^2}{k^2} \frac{\partial f(\omega/k)}{\partial v} \mathrm{,} \label{eq:classical}
\end{equation} 
where $k$ ($\omega$) is the wave vector (frequency), $\omega_{\mathrm{pe}} = (4\pi n_e e^2/m_e)^{1/2}$ is the plasma frequency, and $f$ is the electron distribution function with the normalization condition $\int f d^3 v = 1 $.  The damping rate $\gamma$ can be alternatively derived from the dielectric function formalism:

\begin{equation}
 \frac{\gamma}{\omega}  =\frac{1}{2} \mathrm{Im}\left[\epsilon(\mathbf{k},\omega)\right] \mathrm{,}\label{eq:die}
\end{equation}
where $\epsilon$ is the well-known dielectric function obtained by the random phase approximation.
For high  $k$ values comparable to the electron de Broglie wave length, the dielectric function
should be replaced by the Lindhard function~\cite{Lindhard}: 

\begin{equation} 
\epsilon(\mathbf{k},\omega) = 1 + \frac{\omega_{\mathrm{pe}}^2}{k^2} \frac{m_e}{\hbar} \int 
\frac{ f(E(\mathbf{v}_f)) - f(E(\mathbf{v}_i))}{\omega - (E_f - E_i)/\hbar } d^3 \mathbf{v}\mathrm{,} \label{eq:quantum}
\end{equation} 
where $E(\mathbf{v}_i)$ ($E(\mathbf{v}_f)$) is the electron energy whose momentum is $m_e \mathbf{v}_i = m_e \mathbf{v} - \hbar \mathbf{k}/2$ ($m_e \mathbf{v}_f =m_e \mathbf{v} + \hbar \mathbf{k}/2 $).  The imaginary part of the dielectric function is given from Eq.~(\ref{eq:quantum}) as~\cite{zhu}   

\begin{equation} 
\mathrm{Im}[\epsilon] =\frac{\omega_{\mathrm{pe}}^2}{k^2} \frac{m_e}{\hbar} \int \pi \delta( \omega - \frac{E_f - E_i}{\hbar} )
 \left(f(E_f) - f(E_i)\right) d^3 \mathbf{v}\mathrm{.} \label{eq:imag}
\end{equation}  
In the limit $\hbar \cong 0$,   Eq.~(\ref{eq:imag}) is reduced to Eq.~(\ref{eq:classical}).  
Eq.~(\ref{eq:imag})  describes a process an electron with  $\mathbf{v} = \mathbf{v}_i$ absorbs quanta from the wave  and transitions to the state with $\mathbf{v} = \mathbf{v}_f$ while the total energy ($\hbar \omega = E_f -E_i$) is conserved.
The Landau damping  can be understood more clearly in the co-moving reference frame where the wave is stationary: $ y = x- (\omega/k) t$. 
In this reference frame, the dielectric function $\epsilon^s $ is given as in Eq.~(\ref{eq:quantum}), but the electron distribution becomes shifted by $v_p = \omega/k$ so that $f^s(v) = f(v + \omega/k)$. 
The dielectric functions in the original and the new reference frames are related by $\epsilon(k,\omega) = \epsilon^s(k, \omega-kv_p)$. In particular, the Landau damping of the wave is given as $\gamma/\omega = 1/2~\mathrm{Im}[\epsilon(k,\omega)] = 1/2~\mathrm{Im} [\epsilon^s(k,0)]$; the electrons responsible for the damping have the  initial velocity in the co-moving frame such that $E(v_i + \hbar k/m_e) - E(v_i) = \hbar \omega =0$.

 With a view to estimating the simplest quantum diffraction effect,   consider  a Maxwellian distribution $ f= f_M$   and a weak potential $\phi_0$, where the energy-momentum dispersion relation follows that of the free electron $E(\mathbf{v}) = m_ev^2/2$.  The electron that respects the energy conservation  has  the initial  velocity $v_i =  \hbar k /2m_e $.   The Landau damping is given, from Eqs.~(\ref{eq:die}) and (\ref{eq:imag}), as

\begin{figure}
\scalebox{0.5}{
\includegraphics{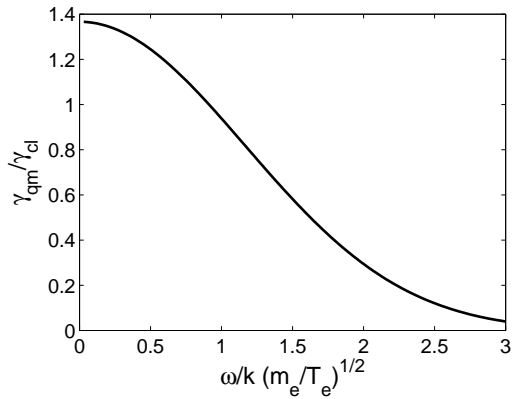}}
\caption{\label{fig:1} The comparison of the damping rates in quantum mechanical
and classical analysis, $\gamma_{\mathrm{qm}}/\gamma_{\mathrm{cl}}$.
$k = 10^8\mathrm{cm}^{-1}$ for a Maxwellian distribution with $T_e = 10 \ \mathrm{eV}$. 
}
\end{figure}

\begin{equation}
  \frac{\gamma_{\mathrm{qm}}}{\omega} =  \frac{\pi}{2} \frac{\omega_{\mathrm{pe}}^2}{k^2}  \frac{m_e}{\hbar} \left(f_M(v_p + \frac{\hbar k}{2m_e}) - f_M(v_p - \frac{\hbar k}{2m_e})\right) \mathrm{,} \label{eq:sim}
\end{equation}
where $v_p = \omega /k $. 
We compare the rate in Eq.~(\ref{eq:sim}) to that in Eq.~(\ref{eq:classical}), when $k = 10^8\mathrm{cm}^{-1}$ for  $T_e = 10 \ \mathrm{eV}$ (Fig.~\ref{fig:1}).  Due to the quantum diffraction effect,  the classical theory  underestimates (overestimates) the damping rate when the wave phase velocity is  lower (higher)  than the electron thermal velocity.

\begin{figure}
\scalebox{0.5}{
\includegraphics{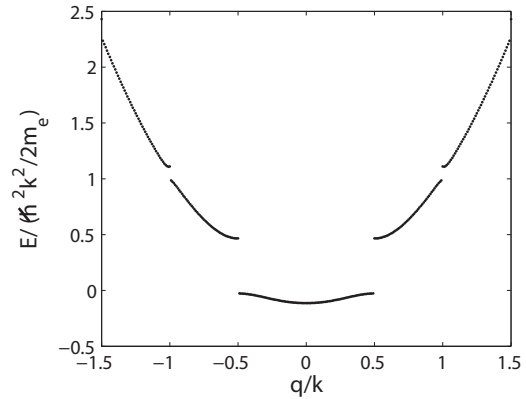}}
\caption{\label{fig:2}  The electron energy $E(q)$ as a function of the wave vector $q$;
$e\phi_0/(\hbar^2k^2/2m_e) =0.5$. 
}
\end{figure}

\begin{figure}
\scalebox{0.5}{
\includegraphics{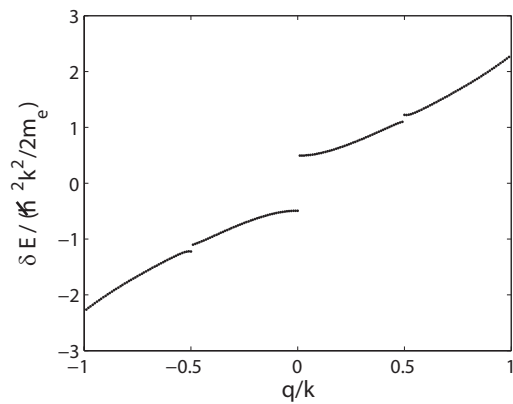}}
\caption{\label{fig:3}  $E_f - E_i$ as a function of the wave vector $q$;
$e\phi_0/(\hbar^2k^2/2m_e) =0.5$.  
}
\end{figure}

\begin{figure}
\scalebox{0.5}{
\includegraphics{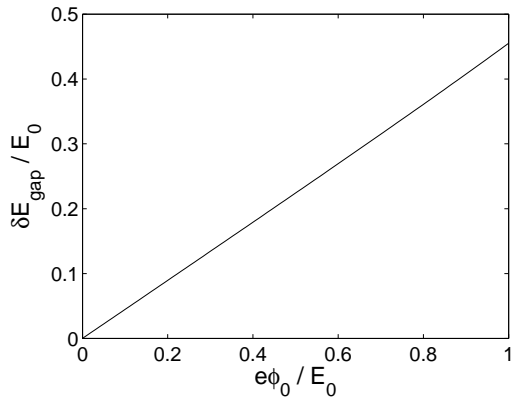}}
\caption{\label{fig:4}
Numerically obtained band gap $\hbar \delta \omega_{\mathrm{gap}}$ as a function of the wave intensity $e\phi_0$. Both axes are scaled by $E_0 = \hbar^2k^2/2m_e$.  
}
\end{figure}

As the intensity of the wave $\phi_0$ increases, 
the energy momentum relation deviates from that of a free electron. 
The energy-momentum relationship of the electrons, $E(q) = \hbar \omega(q)$, where $q$ is  the electron wave vector ($v = \hbar q/m_e$),  is obtained by solving  the Schroedinger's equation in the co-moving frame, 

\begin{equation}
 \hbar \omega(q) \psi = \left( -\frac{\hbar^2}{2m_e} \frac{\partial^2}{\partial x^2}  + e\phi_0 \cos(kx) \right) \psi \label{eq:sch} \mathrm{.}
\end{equation}
The solution can be obtained using the Bloch's theorem $ \psi_{q} = \exp(iqx) u_q(x)$, where $u_q(x)= u_q(x+2\pi/k)$. We note that  Eq.~(\ref{eq:sch}) is the same equation used in the solid state physics for the band gap computation.   
For $e\phi_0 > 0.1 \hbar^2 k^2/2m_e$, an electron band gap appears.
We compute $E(q)=\hbar \omega(q)$ (Fig.~\ref{fig:2}) and
$\delta E(q) = E_f - E_i = \hbar \omega(q+k/2) - \hbar \omega(q-k/2)$ (Fig.~\ref{fig:3})
as a function of $q$ for $e\phi_0/(\hbar^2k^2/2m_e) = 0.5$,
exhibiting a few discontinuous jumps in $\delta E$.
There is no $q$ satisfying $\delta E = 0 $ due to presence of the gap,
hence there is {\em no} Landau damping.
The achievable minimum value of $\delta E = |E_f -E_i|$ is roughly
$\delta E_{\mathrm{min}} \cong \delta E_{\mathrm{gap}} $,
where $\delta E_{\mathrm{gap}} = \hbar \delta \omega_{\mathrm{gap}}$ is the band
gap size when $q \cong 0$.
The band gap size $\hbar \delta \omega_{\mathrm{gap}}$  is computed  as a function of
$e\phi_0/ (\hbar^2k^2/2m_e) $, which shows a linear relationship (Fig.~\ref{fig:4}).
Just like an electron in a solid lattice, which adapts its wave packet to
move freely in the lattice structure instead of being scattered off from the lattice
potential, an electron in the BRS avoids scattering off the pseudo-lattice
formed by the wave.
The quantum diffraction of the electrons is key for this phenomena.

In addition to the band gap, the quantum diffraction also plays a crucial role in the electron trapping, an important process for the frequency detuning, wave breaking and damping~\cite{Oneil,Oneil2}. 
The meaning of ``trapped electrons'' is very different in classical and quantum mechanical pictures.
Classically, an electron is trapped by the wave when the maximum potential is higher than
the electron kinetic energy in the co-moving reference frame.
On the other hand, quantum-mechanically, an electron gets trapped if it is in one of the
bounded states formed by a potential well.
The ground state energy in the Langmuir  wave is given as $1/2 \ \hbar \omega_B$,
where  $\omega_{B} = (e\phi_0/m_e)^{1/2}k$,
assuming the wave minimum is the same as the minimum of the harmonic potential.
The quantum mechanically bounded states exist only when $\hbar \omega_B < e\phi_0$.
Otherwise, there are no bounded states and electrons travel freely across the Langmuir wave. 
If $\hbar \omega_B < e\phi_0$, the size of the wave packet of the bounded ground state is
$k_g = (m_e\omega_B/\hbar)^{1/2}$.  The maximum allowed wave vector  $k_t$  for electrons
to be bounded  is  $k_t = (m_ee\phi_0/\hbar^2)^{1/2}$ so that the minimum wave vector of the electrons $k_g$ is comparable to $k$, unless $\hbar\omega_B \ll  e\phi_0$.  If $k_g$ is comparable to $k$,   electrons behave like quantum waves against the Langmuir  wave and  travel freely. 
The existence of free electrons is partly due to the fact that the wave packet can propagate
freely against the forbidden region via the barrier penetration,  
and partly due to the fact that  the trapped electrons do not lose or gain its average momentum from the wave since the periodicity of the wave does not break the translation symmetry of the electron wave.  This momentum conservation can be explained using  the Bloch's theorem as in the case of the free electrons in metals. If an intense pondermotive potential is suddenly turned on at $t=0$, an electron of an initial momentum $\hbar q$ and an energy $\hbar^2 k^2/2m_e$ would conserve the momentum and the average energy even though the electron wave function branches into different energy bands. Since there is no exchange of the momentum and energy between the wave and the electrons, the effect of electrons on the frequency detuning and damping is negligible.

\begin{figure}
\scalebox{0.5}{
\includegraphics{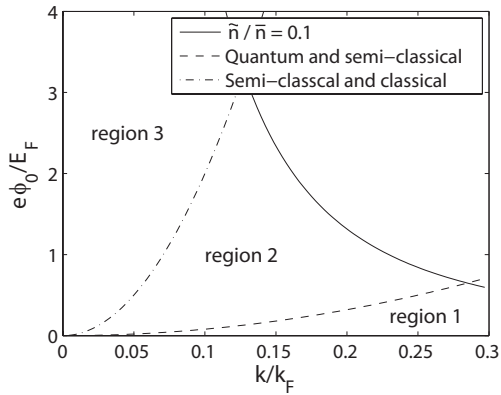}}
\caption{\label{fig:5}  The Regions 1, 2 and 3 (see text for definitions) when $n_e= 10^{24} \mathrm{cm}^{-3}$ and the desirable Langmuir wave intensity $\tilde{n}/\bar{n} = 0.1$.  
}
\end{figure}

For $\hbar \omega_B \ll e\phi_0$,  the ground state wave vector  becomes much bigger than $k$ and the trapped electrons could be described almost classically as in the case of the quantum coherent state approach~\cite{Glauber,Glauber2}.
We do not consider determining the exact boundary for the transition between ``free electrons''  and ``classical bouncing motion''.
We call rather roughly $e\phi_0/\hbar\omega_B<2$  as the Region 1 where the classically trapped electrons are quantum-mechanically  free. 
We choose  $ 2<e\phi_0/ \hbar\omega_B <10 $ as the Region 2 where the  classically trapped electrons  experience both the free electron wave  and the classical  trapping.  In this region, there is some wave breaking due to the classical trapping but the effect is weaker than what is given by the classical theory~\cite{Oneil,Oneil2}. Lastly, we define   $e\phi_0/ \hbar\omega_B >10 $ as the Region 3 where the classical mechanics is valid. A diagram for the the Regions is provided in Fig.~\ref{fig:5}.


We estimate the effect of the quantum diffraction on the Landau damping based on the
above discussion.
If we adopt the literal definition of the delta function in Eq.~(\ref{eq:imag}), 
there cannot be Landau damping since there is no electrons that are resonantly heated by the wave.  For a slightly collisional plasma, we  use  an approximation for the delta function:
\begin{equation} 
\delta ( \omega_f -\omega_i) \cong  \frac{1}{\pi}\frac{\nu_c}{ (\omega_f - \omega_i)^2 +\nu_c^2}  \mathrm{.}
\label{eq:delta}
\end{equation}
where $\omega_{f,i} = E_{f,i}/\hbar$ and $\nu_c$ is the collisional frequency.  
The Landau damping is then reduced by a factor of $\delta \omega_{\mathrm{gap}}/\nu_c$
in comparison to the linear rate given in Eq.~(\ref{eq:imag}).
The electron collision rate in a degenerate dense plasma gets much reduced compared to
the classical prediction~\cite{sonprl}, hence $\delta \omega_{\mathrm{gap}}/\nu_c$
becomes larger.
The appropriate $\nu_c$ would be the electron collision frequency or the electron transit time across the hot spot \cite{Rose}.

According to the classical analysis of the Landau damping~\cite{Oneil2}, if a Langmuir wave reaches a certain intensity, it heats the trapped electrons in the time scale of the bouncing frequency. 
 The distribution of the electrons  then reaches a quasi-steady distribution~\cite{Oneil} and the Landau damping is reduced by a factor of $\omega_{B}/\nu_c$ compared to the linear Landau damping rate. 
Namely, there are two regimes of different dampings:
For $\omega_B < \nu_c$, the damping is given by the linear theory,
and for $\omega_B > \nu_c$, it is reduced by $\nu_c/\omega_{B}$.
On the other hand, in dense plasmas, there are four regimes of different dampings
as previously discussed.
For $e\phi_0/\hbar < \nu_c$, the damping is given by the linear theory. For $\nu_c < e\phi_0/\hbar \leq \omega_B$, it is reduced by $e\phi_0/\hbar\nu_c $ since the gap size is comparable to $e\phi_0$.  In this regime, the wave gets hardly damped even initially.  Especially,  the reduction by the quantum diffraction comes in at much lower intensity of $e\phi_0$ than  that from the formation of the  soliton~\cite{Rose}. 
For $ \nu_c < \omega_B < e\phi_0/\hbar$, the classical and quantum mechanical effects co-exist, and the damping is reduced compared to the linear theory given in Eq.~(\ref{eq:imag}).
  For $\nu_c < \omega_B \ll e\phi_0/\hbar $, it is reduced by $ \nu_c/\omega_B$ as in the classical plasmas. 

 The reduction of the Landau damping in the Region 1 and 2 is beneficial  for  the X-ray BRS compression in dense plasmas. In the BRS, the pondermotive potential $\phi_P$ drives the Langmuir wave $\phi_L$. 
The Langmuir wave intensity is measured by $ \tilde{n}/\bar{n} = (e\phi_L/m_e)(k^2/\omega_{\mathrm{pe}}^2)$,  where $\bar{n}$ is the equilibrium electron density and $\tilde{n}$ is the density modulation in the Langmuir wave.  In the BRS,   it is desirable to have 
\begin{equation}
\frac{\tilde{n}}{\bar{n}} >0.1 \ \mathrm{.}\label{eq:sat}
\end{equation} 
If $k$ is high enough, the Langmuir wave $\phi_L$ would grow to reach $\tilde{n}/\bar{n}>0.1 $ without damping, crossing the Region 1 or the Region 2 in Fig.~\ref{fig:5}. 
For a given pondermotive potential $\phi_P$, the attainable Langmuir wave is $ (1/\epsilon-1)\phi_P$ so that  $\tilde{n}/\bar{n} = (1/\epsilon -1) k^2 e\phi_P/
(m_e\omega_{\mathrm{pe}}^2)$.  The ratio of the maximum attainable Langmuir wave to the prediction from the linear theory is given as $\mathrm{Im} [ \epsilon_l ] / \mathrm{Im} [\epsilon (\phi_0)]$, where  $\epsilon_l$ is the finite temperature Lindhard dielectric function and $\epsilon(\phi_0)$ is the non-linear dielectric function with the band gap structure.  This ratio is very big as computed here.  
As shown in Fig.~\ref{fig:5}, the relevant regime in the BRS compression is $0.1 < k < k_F$, where $k_F= (3\pi^2n_e)^{1/3}$ is the Fermi energy. 
When compressing X-ray pulses, the Landau damping and the wave breaking~\cite{Oneil2}
are very heavy due to the required high electron temperature~\cite{Fisch2}.
However, the heavy damping can be avoided as we have shown here. 

For an example, consider $n_e =10^{25} / \mathrm{cc}$, where $E_F = 169.2  \ \mathrm{eV} $, $\hbar \omega_{\mathrm{pe}} = 117.38 \ \mathrm{eV}$.  For a pump and seed pulse with $k_p \cong 0.1 k_F$, which corresponds to $\omega =2 \times 10^{18} / \sec$ ($\hbar \omega = 1.3 \ \mathrm{keV} $), the pondermotive potential has a wave vector $k = 0.2 k_F$. When $e \phi_0 / \hbar \omega_B = 2$, we can estimate that $\tilde{n} / \bar{n}= \cong 0.05$. If the seed pulse and pump pulse has the same intensity and the pondermotive potential is as strong as  $e \phi_p / \hbar \omega_B = 2$, the eletron quiver energy for each pulse is estimated as $54.14 \ \mathrm{eV} $, which corresponds to the pulse intensity of $3\times 10^{20} \mathrm{W} /  \mathrm{cm}^2 $. The more dense the plasma is, the more strong  the maximum possible intensity of the Langmuir without damping is. 

In this paper, it is shown  that an intense  wave with high wave-vector distorts the energy momentum dispersion relation to generate an electronic band gap which suppresses the Landau damping.
The Landau damping is reduced by a factor of $\delta \omega_{\mathrm{gap}} \tau_c$ compared to the conventional theory, where $\tau_c$ is the mean collision time ($= 1/\nu_c$).
The wave breaking via trapped electrons can be considerably reduced in the Region 1
($e\phi_0/\hbar\omega_B<2$) in Fig.~\ref{fig:5}, where there are no trapped particles.  
The most interesting regime in the BRS compression is $ 0.1 < k/k_F <1 $
where it may even be possible to drive the Langmuir wave to a high intensity without any damping.
Other processes~\cite{Rose,Drake} need to be studied for an alternative saturation mechanism.

While most attention is concentrated  on the aspect of the Landau damping in this paper,
the band gap and the strong modification of the electron wave function  has significant
implications on many other processes in dense plasmas such as the frequency detuning~\cite{Oneil},
the wave breaking~\cite{Oneil2},  the plasmon life time~\cite{Sturm,Sturm2},
the current drive~\cite{fischrev, sonprl},  the BGK mode~\cite{BGK} and wake field accelerator \cite{dodin}.
It would be interesting to see how the overall picture of these physical mechanisms,
as well as recently studied ones~\cite{sonpla}, changes in dense plasmas.

We would like to thank Dr. S. J. Moon for carefully reading the manuscript and for useful discussions and advice.

\bibliographystyle{elsarticle-harv}
\bibliography{<your-bib-database>}

\end{document}